\documentclass[manuscript]{acmart}

\usepackage{booktabs} 
\usepackage[ruled]{algorithm2e} 
\usepackage{tikz}
\usetikzlibrary{arrows}
\usetikzlibrary{shapes.multipart}
\usepackage{caption}

\SetAlFnt{\small}
\SetAlCapFnt{\small}
\SetAlCapNameFnt{\small}
\SetAlCapHSkip{0pt}
\IncMargin{-\parindent}
\usepackage[all]{nowidow}

\usepackage[font=small,labelfont=bf]{caption}
\usepackage[utf8]{inputenc} 
\usepackage[T1]{fontenc}    
\usepackage{hyperref}       
\hypersetup{
    colorlinks=false,
    linkcolor=blue,
    filecolor=magenta,      
    urlcolor=cyan,
    pdfpagemode=FullScreen,
    }
\usepackage{url}            
\usepackage{booktabs}       
\usepackage{amsfonts}       
\usepackage{nicefrac}       
\usepackage{microtype}      
\usepackage{xcolor}         
\usepackage{amsmath}
\usepackage{mathtools}
\usepackage{amsthm}
\usepackage{amsmath}
\usepackage{bbm}
\usepackage{tikz} 
\usepackage{wrapfig}
\usepackage{enumitem}
\usepackage{framed}         

\usepackage[suppress]{color-edits}
\addauthor{ub}{red}
\addauthor{vc}{blue}
\addauthor{jk}{orange}
\addauthor{sw}{blue}
\addauthor{dj}{teal}
\addauthor{mb}{green}
\addauthor{kc}{purple}

\usepackage{caption}
\usepackage{subcaption}

\title{\fls{}: Recording and Incorporating Stakeholder Feedback into Machine Learning Pipelines}

\newcommand{\fl}{\texttt{FeedbackLog}}
\newcommand{\fls}{\texttt{FeedbackLogs}}

\definecolor{pastelGreen}{RGB}{206, 232, 227}
\definecolor{pastelBlue}{RGB}{198, 211, 239}

\usepackage[framemethod=TikZ]{mdframed}
\newenvironment{topbox}
{\begin{mdframed}[roundcorner=5pt, bottomline=false, backgroundcolor=pastelBlue]}
{\end{mdframed}}

\newenvironment{midbox}
{\begin{mdframed}[backgroundcolor=pastelBlue]}
{\end{mdframed}}

\newenvironment{bottombox}
{\begin{mdframed}[roundcorner=5pt, topline=false, backgroundcolor=pastelBlue]}
{\end{mdframed}}

\newenvironment{onebox}
{\begin{mdframed}[roundcorner=5pt,  backgroundcolor=pastelGreen]}
{\end{mdframed}}

\usepackage{array}
\newcolumntype{L}[1]{>{\raggedright\let\newline\\\arraybackslash\hspace{0pt}}m{#1}}
\newcolumntype{C}[1]{>{\centering\let\newline\\\arraybackslash\hspace{0pt}}m{#1}}
\newcolumntype{R}[1]{>{\raggedleft\let\newline\\\arraybackslash\hspace{0pt}}m{#1}}

\setcopyright{none}

\begin{document}

\author{Matthew Barker}
\email{mrlb3@cam.ac.uk}
\affiliation{%
  \institution{University of Cambridge}
  \country{United Kingdom}
}

\author{Emma Kallina}
\affiliation{%
  \institution{University of Cambridge}
  \country{United Kingdom}
}
\affiliation{%
  \institution{Responsible AI Institute}
  \country{United Kingdom}
}

\author{Dhananjay Ashok}
\affiliation{%
  \institution{Carnegie Mellon University}
  \country{USA}
}

\author{Katherine M. Collins}
\affiliation{%
  \institution{University of Cambridge}
  \country{United Kingdom}
}

\author{Ashley Casovan}
\affiliation{%
  \institution{Responsible AI Institute}
  \country{United Kingdom}
}

\author{Adrian Weller}
\affiliation{%
  \institution{University of Cambridge}
  \country{United Kingdom}
}
\affiliation{%
  \institution{The Alan Turing Institute}
  \country{United Kingdom}
}

\author{Ameet Talwalkar}
\affiliation{%
  \institution{Carnegie Mellon University}
  \country{USA}
}

\author{Valerie Chen}
\email{valeriechen@cmu.edu}
\authornote{Both last authors contributed and advised equally. Order was decided by a coin flip. Correspondence to: \href{mailto:valeriechen@cmu.edu}{valeriechen@cmu.edu}  and \href{mailto:usb20@cam.ac.uk}{usb20@cam.ac.uk}}
\affiliation{%
  \institution{Carnegie Mellon University}
  \country{USA}
}

\author{Umang Bhatt}
\email{usb20@cam.ac.uk}
\authornotemark[1]
\affiliation{%
  \institution{University of Cambridge}
  \country{United Kingdom}
}
\affiliation{%
  \institution{The Alan Turing Institute}
  \country{United Kingdom}
}

\renewcommand{\shortauthors}{Barker, et al.}

\begin{abstract}
Even though machine learning (ML) pipelines affect an increasing array of stakeholders, there is little work on how input from stakeholders is recorded and incorporated. We propose \fls{}, addenda to existing documentation of ML pipelines, to track the input of multiple stakeholders. Each log records important details about the feedback collection process, the feedback itself, and how the feedback is used to update the ML pipeline. In this paper, we introduce and formalise a process for collecting a \fl{}. We also provide concrete use cases where \fls{} can be employed as evidence for algorithmic auditing and as a tool to record updates based on stakeholder feedback.
\end{abstract}

\maketitle


\section{Introduction}
\begin{figure}
\centering
\includegraphics[width=0.99\textwidth]{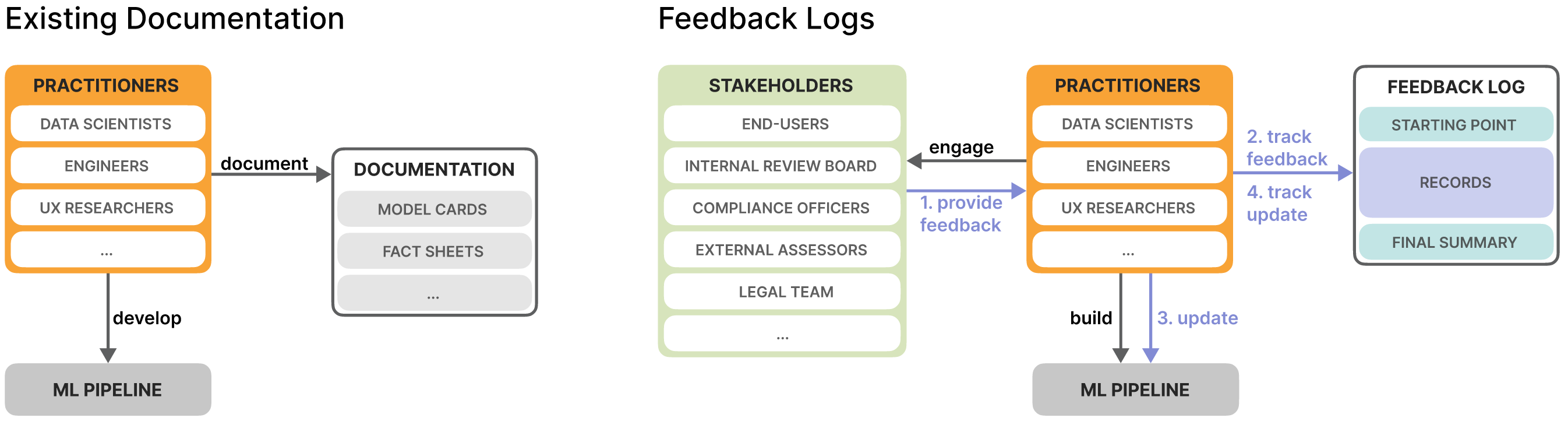}
\caption{(Left) Existing documentation  uses \emph{static} snapshots of a model to document an ML pipeline. 
(Right) In contrast, we propose \fls{} to track the \emph{iterative} development process. Herein, practitioners engage stakeholders for feedback and update the ML pipeline accordingly. While a \fl{} contains a starting point and final summary to bookend stakeholder involvement, the bulk of the \fl{} are the records that document practitioners' \textit{interactions} with stakeholders (shown in purple).
}
\label{fig:summary}
\end{figure}

Stakeholders, who interact with or are affected by machine learning (ML) models, should be involved in the model development process~\citep{fails2003interactive,amershi2014power,cui2021understanding}.
Their unique perspectives, however, may not be adequately accounted for by practitioners, who are responsible for developing and deploying models (e.g., ML engineers, data scientists, UX researchers)~\citep{cheng2021soliciting}. 
We notice a gap in the existing literature around documenting how stakeholder input was collected and incorporated in the ML pipeline, which we define as a model's end-to-end lifecycle, from data collection to model development to system deployment and ongoing usage. 
A lack of documentation can create difficulties when practitioners attempt to justify why certain design decisions were made through the pipeline: this may be important for compiling defensible evidence of compliance to governance practices~\citep{bennett2017information}, anticipating stakeholder needs~\citep{zamenopoulos2007towards}, or participating in the model auditing process~\citep{mokander2021ethics}.
While existing documentation literature (e.g., Model Cards~\citep{mitchell2019model} and FactSheets~\citep{arnold2019factsheets}) focuses on providing \emph{static} snapshots of an ML model, as shown in Figure~\ref{fig:summary} (Left), we propose \fls{}, a systematic way of recording the \emph{iterative} process of collecting and incorporating stakeholder feedback.

The \fl{} is constructed during the development and deployment of the ML pipeline, and updated as necessary throughout the model lifecycle. While the \fl{} contains a starting point and final summary to document the start and end of stakeholder involvement, the core of a \fl{} are the records that document practitioners' interactions with stakeholders.
Each record contains the content of the feedback provided by a particular stakeholder, as well as how it was incorporated into the ML pipeline.  
The process for adding records to a \fl{} is shown in purple in Figure~\ref{fig:summary} (Right). Over time, a \fl{} reflects how the ML pipeline has evolved as a result of these interactions between practitioners and stakeholders. 

To explore how \fls{} would be used in practice, we engaged directly with ML practitioners. Through interviews, we surveyed the perceived practicality of \fls{}. Furthermore, we collected three real-world examples of \fls{} from practitioners across different industries. Each example \fl{} was recorded at a different stage in the ML model development process, demonstrating the flexibility of \fls{} to account for feedback from various stakeholders. The examples show how \fls{} serve as a defensibility mechanism in algorithmic auditing and as a tool for recording updates based on stakeholder feedback.

In summary, the main contributions of this work are:
\begin{enumerate}
    \item A new documentation structure, \fls{}, that captures the iterative process of collecting and incorporating stakeholder feedback (Sections \ref{sec:features_log} and \ref{sec:records}).
    \item Findings from practitioner interviews on the benefits and challenges of implementing \fls{} in practice (Section \ref{sec:practitioner_perspectives}) and an interactive demo tool to make \fls{} more accessible and easy to use for practitioners (Section \ref{sec:demo}).
\end{enumerate}


\section{Overview of \fls{}}
\subsection{Background}
Prior work has focused on documentation that provides a snapshot of the ML pipeline at a specific stage of the ML lifecycle  (Figure~\ref{fig:summary} (Left)). We discuss a few, non-exhaustive, examples of these documentation strategies below. 
Model Cards describe how a model was developed, including who trained the model, when it was trained, and what data was used in the learning procedure along with details of model development and performance of the model on various metrics~\cite{mitchell2019model}. 
Similarly, FactSheets describe relevant information at each phase of the model's development: pre-training, during training, and post-training~\cite{8843893}. 
Explainability Fact Sheets summarize key features that lead a model to be more explainable~\cite{sokol2020explainability}. Reward reports~\cite{gilbert2022reward} frame an ML system as a reinforcement learning model, and record the decisions taken to optimise the system.
Application-specific documentation aims to contextualise more general techniques for use within the domain of interest.  For example, \textit{Healthsheet}~\cite{rostamzadeh2022healthsheet} is a questionnaire adapted from \textit{datasheets for datasets}~\cite{gebru2021datasheets} to improve accountability for data collection and usage in the health domain.  Unlike prior forms of documentation, we propose \fls{} which provide information on the \emph{iterative} process of eliciting and incorporating multiple stakeholder feedback throughout the model's lifecycle (Figure~\ref{fig:summary} (Right)).
To the best of our knowledge, this is the first work that introduces a systematic way to record how stakeholder feedback has been incorporated into an ML pipeline. 
We note that \fls{} can be used alongside existing documentation tools, which we describe further in Section~\ref{sec:incorp}.

The rise of participatory ML~\cite{paml} has resulted in the incorporation of feedback from a diverse set of stakeholders. This raises issues such as ``participation washing''~\cite{sloane2022participation} and a lack of clarity as to what is expected from stakeholders~\cite{birhane2022power}. \fls{} aim to clarify exactly what is expected from stakeholders and the effect of their participation. In addition to documenting model development, previous work has argued for a comprehensive understanding of the usage of a system, including algorithmic auditing~\cite{costanza2022audits,paleyes2022challenges} and critical refusal~\cite{garcia2022no}. By tracking the reasons for decisions prompted by feedback, \fls{} address the \emph{accountability gap}~\cite{raji2020closing} in the development of ML systems that elicit feedback from numerous stakeholders. A \fl{} provides more information than a one-off certification~\cite{henriksen2021situated} and captures the iterative development process rather than a static snapshot~\cite{shergadwala2022human}.



\subsection{\fl{} Components} \label{sec:features_log}
To motivate the design of \fls{}, we set out three desiderata which can be used to evaluate their value added to the documentation process.
\begin{enumerate}
    \item \textit{Completeness:} \fls{} should provide comprehensive details about stakeholder feedback and subsequent practitioner updates. 
    \item \textit{Flexibility:} \fls{} should be able to be integrated into the ML pipeline at any point. \fls{} should also be able to handle the variability in the types and amount of stakeholder feedback as well as the types of updates a practitioner may consider.
    \item \textit{Ease of Use:} \fls{} should come with minimal overhead for practitioners to adopt.
\end{enumerate}

%

\begin{figure}[t]
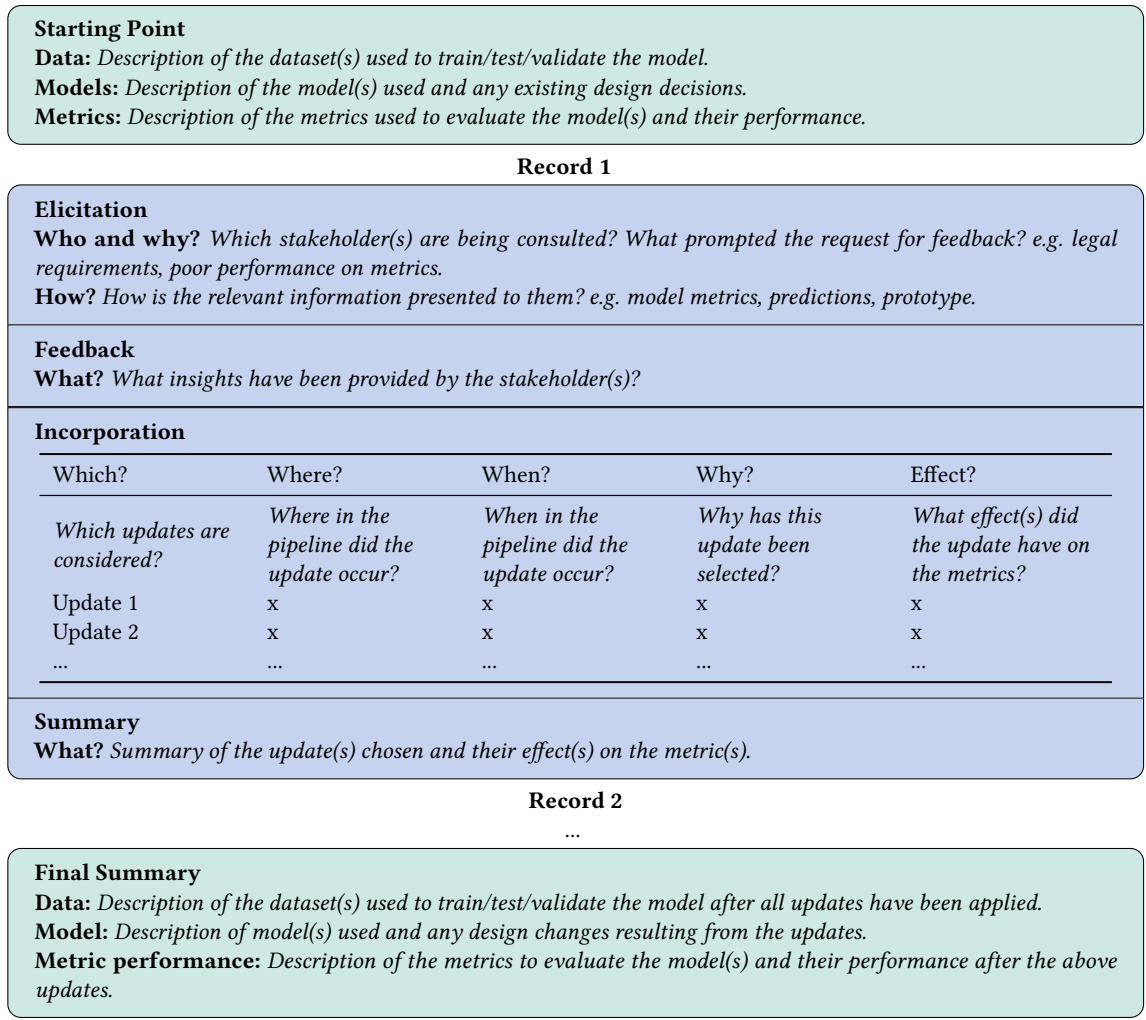

\begin{onebox}
\textbf{Starting Point} \\
\textbf{Data:} \emph{Description of the dataset(s) used to train/test/validate the model.} \\
\textbf{Models:} \emph{Description of the model(s) used and any existing design decisions.} \\
\textbf{Metrics:} \emph{Description of the metrics used to evaluate the model(s) and their performance.}
\end{onebox}

\textbf{Record 1}
\medskip
\begin{topbox}
\textbf{Elicitation} \\
\textbf{Who and why?} \emph{Which stakeholder(s) are being consulted? What prompted the request for feedback? e.g. legal requirements, poor performance on metrics.} \\
\textbf{How?} \emph{How is the relevant information presented to them? e.g. model metrics, predictions, prototype.}
\end{topbox}

\begin{midbox}
\textbf{Feedback} \\
\textbf{What?} \emph{What insights have been provided by the stakeholder(s)?}
\end{midbox}

\begin{midbox}
\textbf{Incorporation}
\begin{center}
\begin{tabular}{L{2.5cm} L{2.5cm} L{2.5cm} L{2.5cm} L{2.5cm}}
\toprule
Which? & Where? & When? & Why? & Effect? \\
\midrule
\emph{Which updates are considered?} & \emph{Where in the pipeline did the update occur?} & \emph{When in the pipeline did the update occur?} & \emph{Why has this update been selected?} & \emph{What effect(s) did the update have on the metrics?} \\
Update 1 & x & x & x & x \\
Update 2 & x & x & x & x \\
... & ... & ... & ... & ... \\
\bottomrule
\end{tabular}
\end{center}
\end{midbox}
\begin{bottombox}
\textbf{Summary} \\
\textbf{What?} \emph{Summary of the update(s) chosen and their effect(s) on the metric(s).}
\end{bottombox}

\textbf{Record 2} \\
...
\medskip

\begin{onebox}
\textbf{Final Summary} \\
\textbf{Data:} \emph{Description of the dataset(s) used to train/test/validate the model after all updates have been applied.}\\
\textbf{Model:} \emph{Description of model(s) used and any design changes resulting from the updates.} \\
\textbf{Metric performance:} \emph{Description of the metrics to evaluate the model(s) and their performance after the above updates.}
\end{onebox}

\caption{The \fl{} includes three sections: a starting point, one or more records, and a final summary. The records section is further divided into the interactions with the stakeholder(s) (elicitation and feedback) and the resulting updates taken by the practitioners (incorporation and summary).}
\label{fig:fl_sections}
\end{figure}

We propose a template-like design for \fls{} with three distinct components (shown in Figure~\ref{fig:fl_sections}): a \textbf{starting point}, one or more \textbf{records}, and a \textbf{final summary}. We describe both the starting point and final summary now and the records in greater detail in the subsequent section. To illustrate how \fls{} can be instantiated, we provide practical examples that have been completed by practitioners in Section~\ref{sec:example_logs}.

The \textbf{starting point} describes the state of the ML pipeline before the practitioner reaches out to any relevant stakeholders. 
The starting point might contain information on the objectives, assumptions, and current plans of the practitioner. More generally, a starting point may consist of descriptions of the data, such as Data Sheets~\cite{gebru2021datasheets}; metrics used to evaluate the models; or policies regarding deployment of the system ~\cite{wan2020human}. 
This component provides \emph{flexibility} since the \fl{} can capture any arbitrary starting point in the development process. 
A proper starting point allows auditors and practitioners to understand when in the development process the gathered feedback was incorporated, and defensibly demonstrates how specific feedback led to changes in the metrics.

The feedback from stakeholders is contained in the \textbf{records} section, which can house multiple records. A single record logs how the stakeholder was requested for feedback, the stakeholder's response, and how the practitioner used the stakeholder input to update the ML pipeline. Figure \ref{fig:fl_sections} shows the structure of one record, which contains the elicitation, feedback, incorporation and summary sections for one source of feedback.
Each record conveys enough information to satisfy \emph{completeness} while not being excessively burdensome to hinder \emph{ease of use}. 

The \textbf{final summary} consists of the same questions as the starting points, i.e. which dataset(s) and models are used after the updates, as well as the metrics used to track model performance. This component provides \emph{completeness} by encapsulating the net effect as a result of feedback from all the relevant experts. 
Proper documentation of the finishing point of the \fl{} allows reviewers to clearly establish how the feedback documented leads to concrete and quantifiable changes within the ML pipeline.

\section{Records}
\label{sec:records}
Each record in a \fl{} is a self-contained interaction between the practitioner and a relevant stakeholder. It consists of how the stakeholder was requested for feedback (\textbf{elicitation}), the stakeholder's response (\textbf{feedback}), and how the practitioner used the stakeholder input to update the ML pipeline (\textbf{incorporation}).

\subsection{Elicitation}
Every record in a \fl{} begins with a practitioner's request for feedback. Tracking how the request was made gives vital context for deciding on how to act on the advice~\cite{stumpf2007toward}, and surface potential downstream issues, such as use of leading prompts, omission of key information, or other problems in the feedback collection process.     

\paragraph{Which stakeholder(s) are being consulted and why?} There are many stakeholders who can provide feedback to improve models~\cite{suresh2021beyond}. 
Stakeholders may be internal to a practitioner's organisation (e.g., senior leadership, compliance officers, account executives) or external (e.g., regulators, auditors, review boards, end users)~\cite{bhatt2020machine,deshpande2022responsible}.
Acknowledging the stakeholder who was consulted is important to document in the feedback procedure, since credit attribution is key to responsible innovation~\cite{solaiman2019release,intelligence2021responsible}. Crediting the source of feedback also helps stakeholders gauge if and when their comments are incorporated into the pipeline~\cite{ouyang2022training, liang2022holistic}. Additionally, it may be important to document \emph{why} the particular stakeholder is being asked for feedback. For example, experts from different fields may be consulted to see whether something noteworthy (e.g., fairness considerations in a specific jurisdiction) has been overlooked. When many stakeholders are consulted for the same reason, as is the case in participatory ML, it is up to the practitioner's discretion whether each stakeholder should be in a separate record, or combined into the same record.


\paragraph{How is the relevant model information presented to stakeholders?} While acquiring stakeholder feedback over a series of interactions~\cite{lee2022evaluatingInteractive}, practitioners will need to decide on what information about a model should be shown to the stakeholder. The information should help the stakeholder develop an appropriate understanding of the current pipeline 
Approaches to communicate such information include socio-technical details~\cite{gebru2021datasheets,mitchell2019model}, performance metrics~\cite{hunton201021st,raji2019actionable}, model explanations~\cite{chen2022interpretable,dodge2019explaining}, and confidence estimates~\cite{bhatt2021uncertainty,kay2016ish}. 
The content and presentation of model information will affect the stakeholder's downstream feedback~\citep{schnabel2018improving}. 

\subsection{Feedback} 
The content of feedback elicited from stakeholders is tracked in each record. Different stakeholders may tend to provide different kinds of feedback, and we illustrate examples below:
\begin{itemize}
    \item \textbf{End Users} are individuals who may be affected by the pipeline. 
    End users can provide feedback on desired model behaviour or feedback on the issues with existing model behaviour. 
    For example, they might specify the kinds of behavior that a model should not exhibit (e.g., a model should not be able to generate hate speech~\cite{brown2020language, markov_2022}). 
    \item \textbf{Regulators} include compliance officers, internal review boards, and independent evaluators. Their feedback may include how to be compliant with regulations~\cite{voigt2017eu,european2020white}, policies~\cite{stafford2018role,cihon2021ai}, or industry standards~\cite{cihon2021ai,nativi2021ai}. These pieces of feedback would need to be translated into concrete actionable updates, which we soon discuss.
    \item \textbf{Domain Experts} are individuals with prior experience and knowledge about the context of the ML pipeline. They may give practitioners auxiliary information that can be used to inform model development (e.g., feature attributions~\cite{weinberger2020learning}, style factors~\cite{adel2018discovering}, semantically-meaningful concepts~\cite{koh2020concept}). 
\end{itemize}

\subsection{Incorporation}
\label{sec:incorp}
Once stakeholders have provided feedback, practitioners can leverage their input to improve the model.
It is imperative to document the update process as there are many different ways (i.e. types of updates) in which a single piece of stakeholder feedback could be incorporated. 
These updates to the ML pipeline can be largely clustered into \emph{model updates} or \emph{ecosystem updates}, which we now describe in more detail. 

\subsubsection{Model updates} It is often feasible to incorporate targeted feedback by making direct changes to the ML model. We focus our discussion on the common supervised learning setting, where a practitioner minimises a loss function on a dataset to learn a model that has many parameters, and any one of these aspects of the model could be changed in response to feedback provided. Common model updates include dataset, loss function, and parameter space updates (a more extensive list can be found in~\citet{chen2023perspectives}): 
\begin{itemize}
    \item \textbf{Dataset updates:} Feedback can be incorporated by adjusting the dataset of a model, i.e. by adding, modifying or removing data~\cite{dao2019kernel,xu2018fairgan,wan2020human}. In addition to active data collection~\cite{irvin2019chexpert}, dataset updates may take place in an unsupervised way~\cite{ratner2017snorkel,hertwig2009description,littlestone1994weighted,hannan1957approximation}.
    
    \item \textbf{Loss function updates.} Feedback can also be used to  update the loss function, thus changing the optimisation objective of the model. It is possible to add constraints to the model which may capture normative notions, such as fairness or transparency~\cite{zafar2017fairness,lakkaraju2016interpretable}, as well as practical considerations, like resourcing or robustness~\cite{frankle2018lottery}.
    
    \item \textbf{Parameter space updates.} Feedback can be incorporated by changing the architecture or features of the model~\cite{correia2019human, roe2020feature}, which affects the model parameter space. These updates traditionally require more technical users, although there are user-friendly interfaces developed to allow even non-technical experts to edit the model in a more direct manner~\cite{wang2021gam,yang2019study}, even in models with many billions of parameters~\cite{meng2022locating, massEdit2022, locateLLMs2023}.

    
\end{itemize}

Implementing such changes to a model requires the practitioner to translate stakeholder feedback into a concrete update, which can be challenging. Not all updates naturally fit in this decomposition. For instance, in large language models \cite{brown2020language, bommasani2021opportunities}, the structure and context of the \textit{prompts} used to elicit generations can have a substantial impact on the model's output~\cite{wei2022chain, zhou2022least, zhou2022large}. Prompts are not necessarily ``data,'' nor parameters; however, their updates are worth tracking nonetheless and naturally fit within the purview of \fl{}.

\subsubsection{Ecosystem updates.} In many practical settings, making model updates only may be insufficient or ineffective to account for a piece of feedback, requiring modifications to the broader ecosystem. Here, ecosystem refers to the socio-technical realm in which the ML pipeline lives. We now describe parts of the ecosystem that can be altered upon receiving feedback.

\begin{itemize}
    \item \noindent\textbf{Documentation.} Feedback can increase the need for documentation. For instance, if the practitioners are made aware of audit requirements (e.g. as outlined in the drafts of the EU AI Act~\cite{EuAIAct} and the Canadian AI and Data Act~\cite{CanadaAIRegulation}), then practitioners might be required to log aspects of the model and its development that have not been considered before. Such aspects could be an additional metric to include in the Model Cards, properties of the dataset that should be in the Datasheet, or a set of specifications that must be reflected in policy documentation. 
    \item \textbf{Interface or UX Updates.} Feedback from end users is essential to ensure a smooth user experience (UX)~\cite{xu2019toward}. Insights into their perception and usability issues with the interface are required to tailor it to their needs. Changes may include considering the perceived trustworthiness of the model~\citep{qin2020understanding}, the required level of interpretability of how the model arrived at a specific decision~\citep{suresh2021beyond}, or even the emotional relationship with a model~\cite{schweitzer2019servant}. These aspects are often addressed via interface changes (e.g. providing forms of explanation~\cite{weitz2019} or recourse~\cite{schnabel2020impact} oranthropomorphizing the model~\cite{zhang2021motivation}). 
   
    \item \textbf{Accountability Structure.} Stakeholders might provide insights into risks that are inherent to a pipeline's use case. Whilst it could be difficult to directly incorporate such feedback into an ML model~\cite{stumpf2007toward}, it might prompt practitioners to identify appropriate strategies to address these risks. For instance, they could establish monitoring processes that detect the manifestation of such risks early on, paired with an action plan with clearly defined responsibilities \cite{CanadaAIRegulation}. This increased awareness would ensure that the practitioners are aware of the risks and their role in preventing potential harms \citep{schiff2020, rakova2021}. 
    
    \item \textbf{Deployment Details.} It may be appropriate to update the intended usage and scope of the pipeline. This includes details of scenarios in which the model is expected to function appropriately, scenarios that should be avoided (e.g. due to data or model drift), or the recommended level of human oversight (and the required expertise of the monitoring individual)~\cite{brundage2018malicious}. This could, for instance, be realized in a guidance document that is issued with the model - similar to a manual - that details the best practices of pipeline implementation and usage, as recommended in \cite{EuAIAct,CanadaAIRegulation}. Such guidance could include where and why pipeline failures may occur with a higher likelihood, how to prevent such failures, what data can and cannot be used in certain circumstances, and generally how to ensure optimal model operation~\cite{shergadwala2022human}. By outlining the context of proper system operation, the operators can quickly establish best practices.
\end{itemize}

Model and ecosystem updates are not necessarily exclusive, since both forms of update may be suitable for a given source of feedback. For example, a practitioner may change both a dataset and loss function, while also adding further details regarding best practices of model use. 
We note that some types of feedback 
(e.g., subjective or qualitative feedback) may be more difficult to translate into updates, which should be noted in the record.
The incorporation section of a record also tracks the following two aspects of the implemented updates:

\paragraph{At which stage of the ML pipeline is the update located?} The feasibility of updates is partly dictated by the current stage of the ML pipeline. Thus, the documentation of where in the pipeline an update is located is part of the justification for the choice of update. Common updates for each of the stages are described further:
\begin{itemize}

    \item \noindent\textbf{Data Collection (pre-training)}:  This is typically when updates are made to the ecosystem or to the dataset (e.g., adding data from underrepresented groups). 
    Other updates might also include feature engineering or model class selection~\cite{turner1999conceptual}.
    \item \noindent\textbf{Model Development (training)}: This includes updates made to the optimisation or learning process of a model (e.g., adding regularization~\cite{zafar2017fairness}, importance weighting~\cite{littlestone1994weighted}, specialized fine tuning~\cite{wortsman2022robust}). 
    \item \noindent\textbf{Model Deployment (post-training)}: Even after the model has been developed, ecosystem-level updates (e.g., interface updates and changes to deployment details) can still occur. We note that the lifecycle of the ML pipeline is not linear; it may be necessary to return to earlier stages and consider their relevant updates.
\end{itemize}

\paragraph{How do we measure the impact of the update?} The final part of this section is a description of how the update(s) affected downstream metrics of interest that were spelled out in the starting point. 
To the extent possible, practitioners should explore performing individual updates, rather than implementing multiple updates simultaneously, to disentangle the isolated effects of the individual updates. 
This measurement can be used in comparing multiple updates to explain the reasoning for selecting from a set of updates, thus demonstrating that other alternatives were considered and ruled out for legitimate reasons. The practitioners may choose to refrain from implementing potential updates, making the justification for inaction in the \fl{} even more important.


\subsection{Summary}
Each record contains a summary of the updates that describes what updates were considered and what their effect was on the metrics of interest. Since each record section may consider multiple \emph{potential} updates, it is important to state which updates are ultimately implemented. 
To enhance readability, the summary should capture the impact of updates, while minimizing the amount of technical detail present about the specific update details.

\section{Towards \fls{} in Practice}
\label{sec:use_in_practice}

We intend to make \fls{} effective for real-world projects. The following section describes three steps that we undertook to bring the \fl{} concept closer to practice as well as to uncover considerations which could affect implementation and usage in real scenarios. First, we collected practitioner perspectives on the concrete implementation of \fls{}. Second, we created an open-source \fl{} generator to make the concept accessible to practitioners, as well as to ease the collection of practitioner feedback. Third, we completed example \fls{} based on consultations with practitioners working on ML pipelines.

\subsection{Practitioner Perspectives \& Future Developments}
\label{sec:practitioner_perspectives}

We conducted semi-structured interviews with three practitioners to gain insight into how \fls{} could be implemented in practice (see Appendix \ref{sec:methods} for the interview guide and details of the method). The responses are summarised below.

\textbf{Responsibilities.} All practitioners expected that a single person would be responsible for the completion of \fls{} for a specific system, i.e. the \textit{\fl{} owner}. This person might be the UX researcher, product manager, analyst, or engineering manager, depending on the type of feedback and development stage. The \fl{} owner would frequently draw on the expertise of other roles to provide input, e.g. on developers to establish the feasibility of technical updates, or the UX designer to propose potential UI solutions. Thus, future versions of \fls{} should have the ability to assign the completion of \fl{} sections to a specific role or person. 

\textbf{Timing of \fl{} Completion.} The timings of when to complete a \fl{} evoked varied responses from the practitioners. For smaller, more confined rounds of feedback collection as in the image recognition example below (Figure~\ref{fig:log_image_recognition}), a post-hoc completion by the analyst was deemed sufficient by a practitioner. However, they agreed that for feedback loops requiring more participating parties, the \fl{} should be filled out alongside the stakeholder involvement process to provide a common point of reference for everyone involved.

\textbf{Expected Benefits of Implementing \fls{}.} The practitioners confirmed many of the benefits of \fls{} mentioned in the previous sections, e.g. the predefined structure that allows for fast information gathering and the benefits regarding audits, accountability, and transparency. The practitioners also suggested that \fls{} might improve communication and knowledge-sharing within organisations. One practitioner mentioned that the product team around the ML model was working with a different information management software than the technical team. They mentioned that this was especially true for A/B tests: the technical team members often had no context around why specific versions were developed and compared, and even lost track of the different versions themselves due to distributed and contradictory information. This resulted in communication issues. \fls{} could serve as a single source of truth that includes links to the other, more specific software. 
Additionally, an interviewee named \fls{} as a repository of past mistakes, solutions, and best practices. If an issue emerged, it could be used to trace the source of the issue as well as to identify past reactions to similar issues and the (long-term) effect of these reactions.

\textbf{Expected Challenges of Implementing \fls{}.} The practitioners anticipated several challenges during the practical implementation of \fls{} that are listed below.

\textit{Log Access.} It is essential to consider who would be able view a \fl{}, amend it, and who would be able to assign these access rights. Since one of the main benefits of \fls{} is that they can increase transparency and accountability, we propose maximum internal viewing access with minimum edit rights. However, this should be customizable to the specific needs of a team. Thus, we plan to incorporate the ability to assign and restrict access in further versions of the \fls{}.

\textit{Scalability: Search and Linking \fls{}}. \fls{} will be created by different \fl{} owners along the entire ML pipeline. Additionally, large organizations often have numerous teams working on various ML models, each of which might require input from many stakeholders. Two practitioners mentioned concerns around organising \fls{} and establishing a structure between the individual entries. 
Future versions of \fls{} could address this concern via the ability to link and search \fl{} entries. 
In many cases, linking \fls{} is essential to trace decisions: For example, initial exploratory user research often scopes product requirements first. These are refined with further user research as well as consultations of the technical team regarding feasibility, both resulting in further \fls{} with more detailed technical requirements. The \fls{} of these different steps should be linked, so it is clear which insights prompted which technical solution. 

\textit{Logistical Trade-offs.} Completing a \fl{} involves a compromise between detail (e.g. the number of different incorporation strategies considered or the level of description of the final update) and labour. Two practitioners mentioned that it might be a nuisance for the \fl{} owner to chase the different required inputs from several team members. However, they agreed that future auditing processes will require detailed process logs for many systems. The current version of \fls{} already offers a high degree of flexibility regarding the depth and detail provided, allowing practitioners to complete it following the depth-labour balance that they deem fit. We plan to maintain and further develop this flexibility in future \fl{} versions.  

\subsubsection{Summary}
The collected practitioner perspectives offered valuable insights into aspects of the \fls{} that could be improved to increase its fit within existing ML pipelines. In addition to the concerns mentioned by the practitioners, we identified three further challenges for practical applications of the \fls{}, given in Appendix \ref{sec:additional_considerations}. To facilitate the collection of stakeholder insights, as well as to make \fls{} accessible for first practical use cases, we introduce an online demo that allows for the quick generation of a \fl{}.

\subsection{FeedbackLog Demo}
\label{sec:demo}
To ease and encourage the adoption of \fls{}, we provide an open-source \fl{} generator\footnote{\url{https://feedback-log.web.app/}}. We acknowledge that this demo is a prototype, solely meant to illustrate the components of \fls{} and to gather feedback on how they may be incorporated into existing workflows. Our tool consists of two components: a web interface for stakeholders, practitioners, and auditors to interact with; and a command-line interface (CLI), shown in Figure \ref{fig:cli}, to enable practitioners to track updates at the source code level\footnote{Code available at \url{https://github.com/barkermrl/feedback-log}}. The \fl{} generator addresses the three desiderata described in Section~\ref{sec:features_log}:
\begin{enumerate}
    \item \textit{Completeness:} The tool covers the spectrum of possible update types: all feedback and ecosystem-level updates are logged in the web interface, while model-level updates are tracked by the CLI. 
    \item \textit{Flexibility:} The web interface is designed to be ecosystem-agnostic, providing a universal interface that can be used alone or with other logging methods. At the time of writing, the CLI only supports Python~\cite{van1995python}.
    \item \textit{Ease of Use:} The web interface contains prompts for expert feedback and structures a \fl{} automatically. To ensure all feedback is incorporated, the CLI has a built-in checklist that consists of the \fl{} components that can be integrated into a practitioner's existing workflow.
\end{enumerate}

In the future, our tool can be extended to a richer interface with which \textit{both} stakeholders and practitioners can interact. This would ease the creation of -- and updates to -- \fls{}, as stakeholders could provide feedback within the tool and practitioners would update the pipeline using our CLI integration. Such a tool would also reduce the burden of maintaining a \fl{}.


\subsection{Example Logs}
\label{sec:example_logs}

\begin{figure}[]
\small
\textbf{Asthma Patient \fl{}}
\medskip
\begin{onebox}
\textbf{Starting Point} \\
\textbf{Data:}  Data of asthma patients, with target as indicator for onset of asthma arrests. \\
\textbf{Models:} Conversational Agent that combines pre-scripted options and model score outputs.  \\
\textbf{Metrics:} No statistical metric yet, objective is to converse with a patient and aid them in managing their conditions
\end{onebox}
\medskip
\begin{topbox}
\textbf{Record 1: Elicitation} \\
\textbf{Who and why?} Clinician. Need clinician insight to understand what details an effective conversational agent should capture. \\
\textbf{How?} Intent of project explained. Clinician was specifically asked to capture all relevant details of an asthma consultation through a mock patient-physician interview. 
\end{topbox}

\begin{midbox}
\textbf{Feedback} \\
\textbf{What?} Received list of questions that clinicians/patients typically ask during clinic sessions. 
\end{midbox}

\begin{midbox}
\textbf{Incorporation}
\begin{center}
\begin{tabular}{L{0.17\linewidth} L{0.17\linewidth} L{0.17\linewidth} L{0.17\linewidth} L{0.17\linewidth}}
\toprule
Which? & Where? & When? & Why? & Effect? \\
\midrule
Add details to metrics & Ecosystem and Metrics & Post Training & Model remains flexible & Model able to provide required details \\
Make dataset of details and fine tune model  & Dataset & Training & Model updated to new information & Unstable training results \\
\bottomrule
\end{tabular}
\end{center}
\end{midbox}
\begin{bottombox}
\textbf{Summary} \\
\textbf{What?} Ecosystem update as part of metrics: added requirements to model to be able to answer certain questions.
\end{bottombox}
\medskip
\begin{topbox}
\textbf{Record 2: Elicitation} \\
\textbf{Who and why?} Clinician. Understanding that the optimal conversational chatbot does not face the same constraints as a clinician and so can ask more detailed questions or spend more time on explanations. \\
\textbf{How?} Clinician asked to explain all information they would like to ask/provide, time-permitting.
\end{topbox}

\begin{midbox}
\textbf{Feedback} \\
\textbf{What?} Clinician provided a list of questions to obtain basic patient information, which can make a significant difference to health outcomes, and does not get communicated during clinician visits because of time limitations. 
\end{midbox}

\begin{midbox}
\textbf{Incorporation}
\begin{center}
\begin{tabular}{L{0.17\linewidth} L{0.17\linewidth} L{0.17\linewidth} L{0.17\linewidth} L{0.17\linewidth}}
\toprule
Which? & Where? & When? & Why? & Effect? \\
\midrule
Include basic details in dataset  & Dataset & Training & Model updated to new information without manual engineering & Model reliably provides complete answers \\
\bottomrule
\end{tabular}
\end{center}
\end{midbox}
\begin{bottombox}
\textbf{Summary} \\
\textbf{What?} Created a base of fundamental information that can be queried and explained  to make it easier for patients to learn the basics of their condition.
\end{bottombox}
\caption{\fl{} for Asthma Conversational Agent project. Creating an AI that can effectively converse with and aid asthma patients requires the domain expertise of  clinicians. The \fl{} shows how a clinician's feedback and its impact on the ML pipeline can be tracked in an organised way.}
\label{fig:log_asthma}
\end{figure}
\begin{figure}[]
\small
\textbf{Image Recognition \fl{}}
\medskip
\begin{onebox}
\textbf{Starting Point} \\
\textbf{Data:} Imagenet1K \cite{imagenet} for training and validation datasets, consisting of 1000 image classes. \\
\textbf{Model:} Convolutional Neural Network (ResNet50 \cite{resnet}). \\
\textbf{Metrics:} None defined yet.
\end{onebox}

\medskip

\begin{topbox}
\textbf{Record 1: Elicitation} \\
\textbf{Who and why?} Hypothetical external assessor vested in the model. Require regulatory approval to use image recognition model in practice. \\
\textbf{How?} Asked for minimum benchmark performance, similar to the 80 percent disparate impact rule~\cite{barocas2016big}.
\end{topbox}
\begin{midbox}
\textbf{Feedback} \\
\textbf{What?} Received a dataset containing adversarial examples of automotive vehicles, along with a minimum accuracy required for this dataset to test the model's robustness.
\end{midbox}
\begin{midbox}
\textbf{Incorporation}
\begin{center}
\scriptsize
\begin{tabular}{L{0.17\linewidth} L{0.17\linewidth} L{0.17\linewidth} L{0.17\linewidth} L{0.17\linewidth}}
\toprule
Which? & Where? & When? & Why? & Effect? \\
\midrule
Imagenet-A \cite{naturaladverserial} with relevant automotive classes & Dataset & Pre-Training & Tests model robustness & Testing dataset for model \\
Minimum accuracy $>50\%$  & Ecosystem \& Metrics & Training & Required for regulatory approval & Benchmark when testing model\\
\bottomrule
\end{tabular}
\end{center}
\end{midbox}
\begin{bottombox}
\textbf{Summary} \\
\textbf{What?} Dataset update: provided new dataset to test the model's robustness when recognising automotive vehicles. Ecosystem update as part of metrics: added requirement that model should achieve $>50\%$ accuracy (robustness) on test dataset.
\end{bottombox}

\medskip

\begin{topbox}
\textbf{Record 2: Elicitation} \\
\textbf{Who and why?} Hypothetical compliance team. Need to ensure model meets external requirements set by industry regulators, as well as internal company policies.\\
\textbf{How?} Presented with current performance on testing dataset recommended by regulator, along with example predictions.
\end{topbox}

\begin{midbox}
\textbf{Feedback} \\
\textbf{What?} Current robustness (34\%) isn't sufficient to meet requirements. In addition, the model is overconfident in its predictions which may cause serious accidents that are unacceptable under company policy.
\end{midbox}

\begin{midbox}
\textbf{Incorporation}
\begin{center}
\scriptsize
\begin{tabular}{L{0.17\linewidth} L{0.17\linewidth} L{0.17\linewidth} L{0.17\linewidth} L{0.17\linewidth}}
\toprule
Which? & Where? & When? & Why? & Effect? \\
\midrule
ResNet-101 \cite{resnet} & Parameter space & Before training & Identify complex features  & Robustness 39\% \\
MEAL V2 \cite{meal} & Loss function & During training & Soften labels & Robustness: 47\% \\
CutMix \cite{cutmix}& Dataset & Before training & Background invariance & Robustness: 48\% \\
\bottomrule
\end{tabular}
\end{center}
\end{midbox}
\begin{bottombox}
\textbf{Summary} \\
\textbf{What?} Used ResNet-101 model with CutMix for data augmentation, since when both updates are used the robustness is 55\%, which exceeds the minimum requirement of 50\%.
\end{bottombox}
\medskip
\begin{onebox}
\textbf{Final Summary} \\
\textbf{Data:} Imagenet1K \cite{imagenet} augmented with CutMix for training, Imagenet-A \cite{naturaladverserial} with relevant automotive classes for testing. \\
\textbf{Model:} Convolutional Neural Network (ResNet-101 \cite{resnet}). \\
\textbf{Metric performance:} 55\% robustness on Imagenet-A testing dataset.
\end{onebox}

\caption{\fl{} for a hypothetical Image Recognition Task. Developing a model that can be used in automative vehicles requires the approval of regulators. Record 1 shows how relevant requirements can be obtained from an external regulator, and Record 2 shows how multiple updates can be combined to meet these requirements. All statistics reported are real values computed by the authors.}
\label{fig:log_image_recognition}
\end{figure}

We now walk through concrete examples of \fls{}: three \fls{} obtained from industry practitioners on ML pipelines  still in development and one demonstration log using a real dataset and model that shows a completed pipeline.

\subsubsection{FeedbackLogs in Industry.}

We collected \fls{} from three practitioners for ML pipelines that they are working on or have recently worked on. They were provided with a blank \fl{} template that they completed in their own time. More details on the methods can be viewed in Appendix \ref{sec:methods}. 
Since the practitioners chose ongoing projects, we refrain from providing the Final Summary section. 
Additionally, to avoid sharing specific information about proprietary ML models, these \fls{} focused on higher level pieces of feedback that practitioners have encountered. 
As such, more complete \fls{} in practice may be much lengthier or messier than the examples that we provide.
We describe the projects and the learnings from each \fl{}.


\smallbreak
\noindent\textit{Asthma Conversation Agent:} This \fl{} describes the project of a national healthcare body to develop a conversational agent for asthma patients, operating via WhatsApp. The aim is to help patients with the management of their condition, including the prediction of the onset of asthma attacks. The \fl{} (Figure~\ref{fig:log_asthma}) contains two records that demonstrate how practitioners can track the needs of stakeholders. At the starting point, there was no statistical metric was defined by the practitioners; however, the log provides evidence that the metrics  eventually used in the project are informed by consultations with clinicians, who are domain experts on asthma. In case of an audit, practitioners can demonstrate how alternate methods of incorporating the feedback were considered, herein adding details to metrics or fine-tuning the model. The summary captures why a particular update (i.e., metric details) was selected.

\smallbreak
\noindent\textit{Recommender Systems:} Next, the \fl{} describes a model developed by a large streaming platform, aiming at increasing the user engagement of subscribed users. The \fl{} (Figure~\ref{fig:log_reco}) shows how the structure of the log is capable of capturing end-user needs and translating this feedback into a concrete UX update. This update manifests as the addition of a ``like'' button to gauge user preferences over repeated interaction, and improves the click-through rate metric used to measure performance for this application.

\smallbreak
\noindent\textit{Sexual Health:} This \fl{} concerns the healthcare domain, focusing on sexual health. A national healthcare provider developed a model to automatically offer treatment for patients suffering from chlamydia symptoms, based on their answers to a questionnaire. The aim of the described stakeholder involvement was to identify accessibility and usability issues for vulnerable demographic groups, risking inaccurate treatment allocation. While both the previous records document feedback provided to projects where the ML pipelines are already set up, this \fl{} (Figure~\ref{fig:log_std}) captures changes that occur in the data collection phase before a model is even trained. The feedback collected from patients and psychologists informed  practitioners that their dataset collection must better accommodate individuals from vulnerable demographic groups. This log could be used as evidence to demonstrate how the organisation took into account the conditions of  vulnerable patients, who now have an alternative method for having their data collected in a way that minimises the risk of unrepresentative data. 
\smallbreak
The example \fls{} provided useful insights in the template's ability to represent the feedback collection and model updating process. The \fls{} concisely tracked the incorporation of feedback for each project, showcasing the flexibility of the \fl{} to describe changes to the pipeline at various stages. 

\subsubsection{Demonstration of a Complete FeedbackLog.}

While the three industry examples demonstrate how \fls{} can be used in the real-world, industry practitioners are prevented from sharing proprietary information about the exact models that are being developed.
As such, we provide a demonstration of a \textit{complete} \fl{}, which uses a real dataset and model, and includes details of technical updates. We consider a hypothetical scenario wherein a practitioner is developing an image recognition model for automotive vehicles.

\smallbreak
\noindent\textit{Image Recognition:} This \fl{} (Figure~\ref{fig:log_image_recognition}) shows records that track non-technical, ecosystem updates as well as technical, model updates. In this case, two updates (to the parameter space and dataset) needed to be used simultaneously since no individual update was sufficient to meet the metric requirements. 
However, we note that individual updates are still tracked.
This \fl{} contains a final summary, as the updates per the second record satisfy specified metrics. 

\section{Conclusion}
Stakeholder engagement is important to consider when deploying ML pipelines. However, even when stakeholders are consulted by practitioners, their feedback is rarely tracked and incorporated in a systematic manner. 
In this work, we propose \fls{}: a tool for practitioners to document the process of collecting and incorporating stakeholder feedback into the ML pipeline. 
\fls{} are designed to be complete, flexible, and easy to use.
Through real-world examples, we demonstrate how \fls{} can record a wide variety of stakeholder feedback and capture the resulting updates made to ML pipelines. 
We hope \fls{} usher in the development of extensible tools for practitioners to empower the voice of a diverse set of stakeholders.

\bibliographystyle{ACM-Reference-Format}
\bibliography{acmart}

\newpage
\appendix

\section{Additional Example Logs}
\begin{figure}[h]
\textbf{TV Content Recommendation \fl{}}
\medskip
\begin{onebox}
\textbf{Starting Point} \\
\textbf{Data:} User watch history of TV content (films, TV series, sports events), including time of day something was watched and the percentage of the content that was watched. User details, such as time being subscribed to the service.\\
\textbf{Models:} Model uses a Convolutional Neural Network architecture to provide personalised recommendations for what to watch next.\\
\textbf{Metrics:} Click-through rates of top-N provided recommendations and the watch percentage of the recommended content.
\end{onebox}
\textbf{Record 1}
\medskip
\begin{topbox}
\textbf{Elicitation} \\
\textbf{Who and why?} Machine Learning Engineers and Data Scientists. Poor performance on metrics, suggesting users were not watching the content that was recommended.\\
\textbf{How?} Model metrics in the form of click through rates and watch percentages, and a test interface to view what content gets recommended based on different watch histories.
\end{topbox}

\begin{midbox}
\textbf{Feedback} \\
\textbf{What?} The evaluation metrics used were not adequate. If someone had watched what was recommended to them, but only watched the first half, this would be deemed a successful recommendation. Yet there is no indication that the user appreciated the recommendation or enjoyed the content. 
\end{midbox}

\begin{midbox}
\textbf{Incorporation}
\begin{center}
\begin{tabular}{L{0.17\linewidth} L{0.17\linewidth} L{0.17\linewidth} L{0.17\linewidth} L{0.17\linewidth}}
\toprule
Which? & Where? & When? & Why? & Effect? \\
\midrule
“Like” buttons were added to content recommendations & Ecosystem and Dataset  & Post Training and Deployment & To provide user supervision for the recommendation algorithm & Increased user click through rate and watch time\\
\bottomrule
\end{tabular}
\end{center}
\end{midbox}
\begin{bottombox}
\textbf{Summary} \\
\textbf{What?} In order to better discern whether a particular recommendation was effective, the users of the TV service were given the ability to “Like” or “Dislike” particular content. This feedback was incorporated as a feature to the machine learning model, which tailored the machine learning model according to what the user liked and disliked. Users started to watch the recommended content more than they did previously.
\end{bottombox}
\caption{\fl{} for TV content recommendation. The \fl{} shows how technical system design choices can be represented in the \fl{} records. }
\label{fig:log_reco}
\end{figure}

\newpage
\begin{figure}[h]
\textbf{Sexual Health \fl{}}
\medskip
\begin{onebox}
\textbf{Starting Point} \\
\textbf{Data:}  NHS standards provide mandatory questions which determine whether treatment can be given or not. For example, certain treatments for forbidden when patients are pregnant, or suffer from certain allergies. 
\\ \emph{Qualitative data:} user requirements from generative user research, including those who are gender diverse or may have a learning disability.
\\ \emph{Quantitative data:} number of misunderstandings from each questions/answer option during user testing.
\\
\textbf{Models:} Model objective is to decide which chlamydia positive patients are eligible to get chlamydia treatment. Through a range of online questions with multiple choice answer options (personalised risk assessment), it checks on medication use, allergies, and other variables that could impact the decision.\\
\textbf{Metrics:} 
\begin{itemize}
	\item Randomized Control Trial will be conducted across the UK to measure safety and effectiveness in comparison to regular offline care.
	\item Number of true positives and negatives, false positives and negatives of prescriptions that the model suggested, compared to offline prescriptions by a clinician.
	\item Time between testing positive and receiving treatment.
	\item Number of people receiving treatment.
	\item Time clinicians spent on patient (offline and online support). 
\end{itemize}
\end{onebox}
\textbf{Record 1}
\medskip
\begin{topbox}
\textbf{Elicitation} \\
\textbf{Who and why?} Patients with chlamydia, sexual health clinicians and health psychologists. Legal requirements, safety risks (e.g. prescribing this treatment to people who are allergic/pregnant can severely impact health).\\
\textbf{How?} Exploratory but also visually (at a later stage showing the questions and decision tree) to capture all relevant details and questions that needed to be included in the risk assessment/online consultation.\\
\end{topbox}

\begin{midbox}
\textbf{Feedback} \\
\textbf{What?} For vulnerable demographic groups, unaccompanied online consultations can be dangerous because the patient might not interpret the question correctly and may not know the information requested. 
\end{midbox}

\begin{midbox}
\textbf{Incorporation}
\begin{center}
\begin{tabular}{L{0.17\linewidth} L{0.17\linewidth} L{0.17\linewidth} L{0.17\linewidth} L{0.17\linewidth}}
\toprule
Which? & Where? & When? & Why? & Effect? \\
\midrule
'I don't know' option was added to applicable questions & Forms and Decision Charts & Early Development Stage & To ensure individuals who are unsure do not select a misleading option & To be measured, but expect fewer false negatives\\
\bottomrule
\end{tabular}
\end{center}
\end{midbox}
\begin{bottombox}
\textbf{Summary} \\
\textbf{What?} This prompted the inclusion of a ‘don’t know’ answer option which would lead to a help screen. In some cases, users are asked to call the helpline and discuss with a sexual health expert, allowing the interview process to proceed in a way that is more reliable. 
\end{bottombox}
\caption{Feedback log for sexual health case study: often feedback is needed even before a dataset is created or a model pipeline exists. The \fl{} is capable of representing and documenting decisions made even at early stages of development. }
\label{fig:log_std}
\end{figure}

\newpage
\section{Command Line Interface (CLI) Usage}
\begin{figure}[h]
    \centering
    \textbf{View \fl{}}
    \medbreak
    \includegraphics[width=\linewidth]{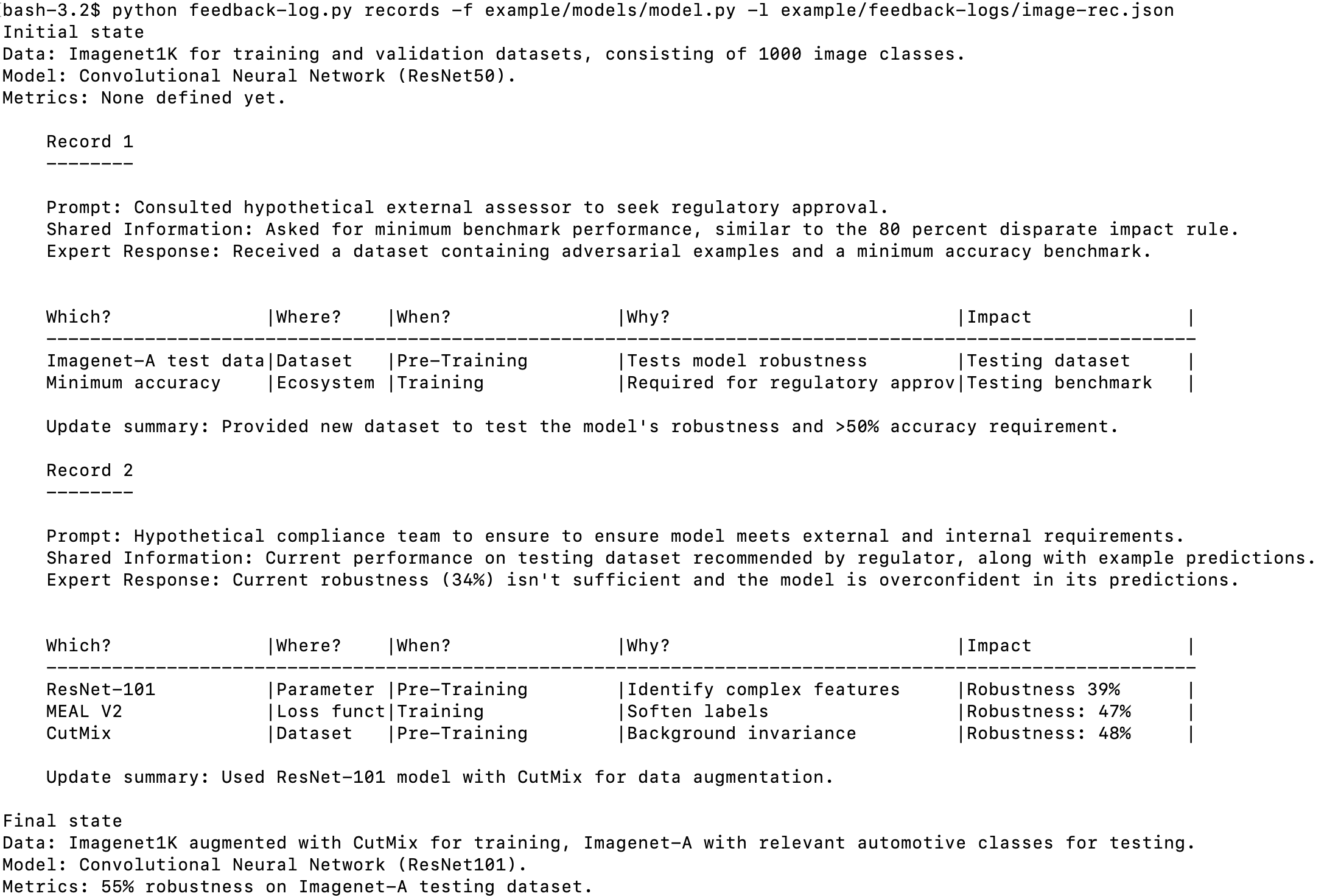}
    \medbreak
    \textbf{View Feedback to be Incorporated}
    \medbreak
    \includegraphics[width=\linewidth]{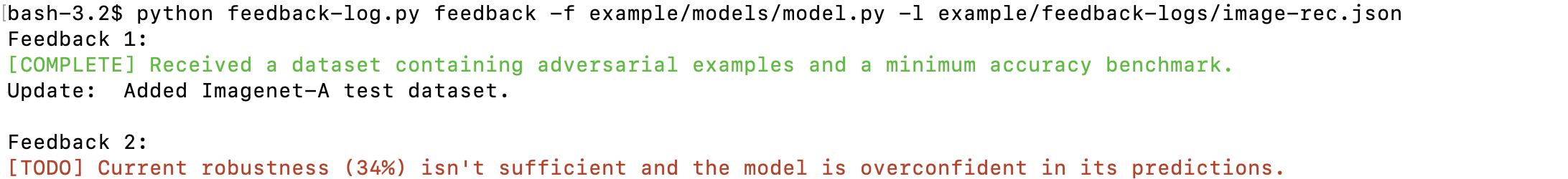}
    \caption{Screenshots showing how the CLI can be used to help practitioners view the \fl{} (Top) and incorporate the feedback from stakeholders (Bottom). The CLI reads the source code comments and automatically updates the checklist (Bottom) once practitioners make updates.}
    \label{fig:cli}
\end{figure}

\newpage
\section{Practitioner Engagement: Methods}
\label{sec:methods}
The following section provides details to the method of the practitioner engagement steps described in section \ref{sec:use_in_practice}, i.e. the semi-structured interviews (section \ref{sec:practitioner_perspectives}) and the example \fls{} (section \ref{sec:example_logs}). Both steps were approved by the Ethics Committee of the University of Cambridge.

\subsection{Semi-Structured Interviews}
The ML practitioners for the semi-structured interviews were recruited via the personal or professional networks of the researchers. 
Each interview lasted between 45 and 60 minutes and was conducted via a video call. The three practitioners had different roles along the ML pipeline, i.e. UX researcher, developer, and engineering manager with varying levels of experience (from one year to over five years). 
The interviews followed an interview guide in a semi-structured manner. The guide included sections on (1) the practitioners' experience and role within ML, (2) their awareness and practices around current stakeholder involvement and their perception of this, (3) high-level feedback regarding the idea and usefulness of \fls{}, and lastly (4), feedback on the timings, responsibilities, and challenges they foresee when applying \fls{} in a specific scenario. There was time for the participants to ask questions and add additional thoughts. The fourth section was supported with a Miro board that displayed an empty \fl{} template. This template was used to discuss the order in which the different sections would be completed in practice, the responsibilities for completing the different sections, and the agency of the different roles in determining the content of these sections. 
The interviews were recorded, summarised, and analysed using thematic analysis~\cite{braun2012thematic}.

\subsection{Example FeedbackLogs}
As for the semi-structured interviews, practitioners that were consulted for the example \fls{} were recruited via personal and professional networks. Two practitioners worked as UX researcher and designer, the third practitioner was a developer. They had between three and nine years of experience in their role. The practitioners were provided with an online document that included the sections of the \fls{} as headers with a short description of the content that such a section would entail. Then, they were asked to complete each section for an ML project they are working on or have recently worked on. This could be done asynchronously in their own  time. The completed documents are the core of the example \fls{}, with slight edits and cuts to increase conciseness. 

\section{Additional Considerations}
\label{sec:additional_considerations}
In addition to the concerns mentioned by the practitioners, we identified three further challenges for practical applications of the \fls{}.


\textit{Measurability of Impact.} Assessing the impact of an update implemented in response to stakeholder feedback can be challenging. Some updates have effects which are hard to define empirically, such as trust or accessibility. In such cases, practitioners could consider expanding the tracked metrics to give a more holistic picture of the pipeline and its objectives.


\textit{Reproducibility.} If third parties rely on \fls{} to reproduce models and replicate a development process, it is essential that practitioners meticulously create and maintain their \fls{} with sufficient detail. For some pipelines, this may include the need to track which how much of the pipeline was procured from third-party vendors.
For instance, if a practitioner fine-tunes a procured large language model \cite{brown2020language, bommasani2021opportunities} for a particular task, they should denote this in the \fl{} but also request thorough documentation of the base model.

\textit{Privacy.} Logging stakeholder feedback may make it possible to identify the stakeholder who provided the feedback. Care should be taken to ensure that stakeholders who may be identifiable have explicitly consented. Stakeholders  who have not consented should have their feedback recorded in a way that does not compromise their anonymity.

\end{document}